\documentclass[preprint,pra,amssymb,amsmath,showpacs,floatfix]{revtex4}

\usepackage{graphicx}
\usepackage{amssymb}
\usepackage{amsmath}

\newcommand{\be}{\begin{equation}}
\newcommand{\ee}{\end{equation}}
\newcommand{\bea}{\begin{eqnarray}}
\newcommand{\eea}{\end{eqnarray}}

\begin{document}

\title{ Comparison of Classical Chaos with Quantum Chaos }

\author{L.A.~Caron$^{a}$, D.~Huard$^{a}$, H.~Kr\"{o}ger$^{a}$$\footnote{Corresponding author, Email: hkroger@phy.ulaval.ca}$,  
G.~Melkonyan$^{a}$, K.J.M.~ Moriarty$^{b}$ 
and L.P.~Nadeau$^{a}$}

\affiliation{
$^{a}$ {\small\sl D\'{e}partement de Physique, Universit\'{e} Laval, Qu\'{e}bec, Qu\'{e}bec G1K 7P4, Canada} \\ 
$^{b}$ {\small\sl Department of Mathematics, Statistics 
and Computer Science, Dalhousie University, Halifax N.S. B3H 3J5, Canada} 
}

\begin{abstract}
We investigate chaotic behavior in a 2-D Hamiltonian system - oscillators with anharmonic coupling. We compare the classical system with quantum system. Via the quantum action, we construct Poincar\'e sections and compute Lyapunov exponents for the quantum system. We find that the quantum system is globally less chaotic than the classical system. We also observe with increasing energy the distribution of Lyapunov exponts approaching a Gaussian with a strong correlation between its mean value and energy.  
\end{abstract}

\pacs{03.65.-w, 05.45.Mt} 

\maketitle 
      


\section{Introduction}
\label{sec:Intro}
Quantum chaos, which denotes properties of quantum systems whose underlying classical system is chaotic, has been experimentally observed in irregular 
energy spectra of nuclei, of atoms perturbed by strong electromagnetic fields \cite{Friedrich}. In a weak exterior magnetic field, however, a hydrogen atom shows a level distribution being neither Poissonian nor Wignerian. How can we then compare classical chaos with quantum chaos? Is the quantum system more or less chaotic than the corresponding classical system? 
An understanding of how classically regular and chaotic phase space is reflected in quantum systems is an open problem, since semiclassical methods of quantisation (EKB, Gutzwiller's trace formula) are not amenable to mixed dynamical systems \cite{Bohigas2}. In order to describe nearest neighbor energy level spacing duistributions for mixed systems, workers have used different approaches. 
Percival \cite{Percival73} has suggested a distribution based on a division of phase space into regular and irregular regions.
Brody \cite{Brody81} has suggested a phenomenological fit between a Wigner and a Poisson distribution. Izrailev \cite{Izrailev90} suggested a distribution based on the concept of quantum localisation. Another interpolation between 
ensembles has been proposed by Lenz and Haake \cite{Lenz91}. The interpolating distribution proposed by Berry and Robnik \cite{Berry84} uses an interpolation parameter determined from classical phase space.

In order to address the above questions, workers have suggested concepts to reintroduce trajectories into quantum mechanics. One possibility is the use of the standard effective action, which takes into account quantum corrections of the classical action. Cametti et al. \cite{Cametti} have considered the model of the 2-D anharmonic oscillator and computed the effective action via loop ($\hbar$) expension to low order. 
Another approach is Bohm's interpretation of quantum mechanics which expresses the Schr\"odinger equation in polar form. It allows to introduce the concept of a particle trajectory and hence a quantum equivalent phase space. This approach has been used by Schwengelbeck and Faisal \cite{Schwengelbeck} to describe chaos in the kicked rotor and later by other workers for the 2-D anisotropic harmonic oscillators \cite{Parmenter}, the hydrogen atom in an external electromagnetic field \cite{Iacomelli} and coupled anharmonic oscillators \cite{Partovi}.
For the kicked rotor in 1-D, which is classically chaotic, it turns out that the quantum system is nonchaotic, i.e. the quantum Kolmogorov-Sinai entropy goes to zero \cite{Schwengelbeck} (rediscovered in \cite{Partovi}). In contrast to that,
the 2-D anisotropic harmonic oscillator, being classically non-chaotic, has been reported to exhibit quantum chaos \cite{Parmenter}.  

A comparison with respect to chaotic behavior of the classical system and the quantum system has been investigated also in field theoretic models. Casetti et al. \cite{Casetti} considered the large $N$ limit of $N$-component $\phi^{4}$ oscillators in the presence of an external field. 
Using mean field theory they observed a strong suppression of quantum chaos due to quantum corrections moving the system away from a hyperbolic fixed point responsable for classical chaos.
Matinyan and M\"uller \cite{Matinyan} considered the model of massless scalar electrodynamics, which is classically chaotic. Using effective field theory and loop expansion, they noticed that quantum corrections increase the threshold for chaos by modifying the ground state of the system.

To address the above questions, we apply the concept of the quantum action, developed in Refs.\cite{Q1,Q2,Q3,Q4,Q5,Q6}. It expresses quantum transition amplitudes $G(x_{fi},t_{fi};x_{in},t_{in})$ in terms of a local action - the quantum action - $\tilde{S}[x]$ defined by 
\begin{eqnarray}
\label{eq:DefQAction}
&& G(\vec{x}_{fi},t=T;\vec{x}_{in},t=0) = \sum_{trajectories} \tilde{Z} \exp[ i \tilde{S}[\tilde{\vec{x}}_{traj}] /\hbar ] ~ ,
\nonumber \\
&& \tilde{S}[\vec{x}] = \int_{0}^{T} dt ~ \frac{1}{2} \tilde{m} \dot{\vec{x}}^{2} - \tilde{V}(\vec{x}(t)) ~ ,
\nonumber \\
&& \tilde{\vec{x}}_{traj} : \delta \tilde{S}[\tilde{\vec{x}}_{traj}] = 0 ~ .
\end{eqnarray}
Here $\tilde{S}$ is evaluated along its trajectory (stationary point of $\tilde{S}$) going from boundary point $(\vec{x}_{in},t=0)$ to $(\vec{x}_{fi},t=T)$. 
There may be several such stationary points. 
The quantum action has been explored  numerically \cite{Q1,Q2,Q3,Q6} for confinement-type potentials (bound state spectrum), e.g. the inverse square potential and polynomial potentials. The numerical studies taking into account only one trajectory (of lowest action) show that the parameters of the quantum action vary smoothly with transition time $T$, interpolating between the classical action at $T=0$ and some "deep" quantum action in the asymptotic regime $T \to \infty$. In the limit of large imaginary transition time, the quantum action has been proven to exist, and to give an exact parametrisation of transition amplitudes \cite{Q4}. Also in this limit analytical relations exist between the the classical potential, the quantum potential and wave functions \cite{Q5}. 

What is the physical interpretation of the quantum action and its trajectories? In the path integral representation of the transition amplitude occurs the classical action and all kinds of paths (zig-zag curves). In the quantum action representation occurs the quantum action but only some trajectories (differentiable curves). Thus we interpret the quantum action as a renormalisation effect on the classical action. Quantum fluctuations appear as tuned parameters in the new action. This situation is similar to renormalisation in solid state physics: A charged particle propagating in the solid interacts with atoms. This results in an effective mass and charge, different from that of free propagation. 

The local quantum action and its trajectories allows to construct a quantum analogue phase space. From that one can compute quantum Poincar\'e sections,  Lyapunov exponents or the Kolmogorov-Sinai entropy like in classical nonlinear dynamics. This gives a new vista on quantum chaos. First results of a numerical study have been reported in Ref. \cite{Q3}. The quantum action has been found useful also to characterize quantum instantons \cite{Q2}. 
The definition of the quantum action looks conspicously similar to that of Bohm's parametrisation of the wave function.
So what is the difference and what is new here compared to previous studies of quantum chaos based on Bohm's approach? 
First, in Bohm's approch considers the quantum system in a state given by a particular wave function. Hence the fluid dynamical motion depends on such wave function (via Bohm's quantum potential). The quantum action, however, is the same for all initial and final states of the transition amplitude, and depends only on transition time. Of course, the trajectories of the quantum action depend on initial and final boundary conditions. 
Second, in Bohm's approach, the mass of the particle occuring in the fluid dynamics picture is the same mass as in the original Hamiltonian. In the quantum action, the mass is renormalized and in general differs in value from the classical mass.
Finally, in contrast to Bohm's approach, the quantum action is very close in functional form to the classical action, and hence allows a direct and detailed comparison of phase space, as will be shown below.

\section{Existence amd analytical form of quantum action in the limit of large imaginary time}
\label{sec:Existence}
The proof of existence of the quantum action in the limit of large imaginary time for Hamiltonian systems has been given for 1-D in Ref.\cite{Q4}. Because we consider in this work a 2-D Hamiltonian system, and for sake of self consistency we will outline the proof here for arbitrary dimension (in particular $D=2,3$). Moreover, we present nonlinear differential equations relating the classical potential to the quantum potential. Also the relation between the quantum action and supersymmetric quantum mechanics will be discussed.
Let us go over to imaginary time $t \to -it$. Then the transition amplitude becomes the Euclidean transition amplitude
\begin{equation}
\label{eq:EuclTransAmpl}
G_{E}(\vec{x}_{fi},T; \vec{x}_{in},0) 
= \langle \vec{x}_{fi} | e^{- H T/\hbar} | \vec{x}_{in} \rangle 
= \left. \int[dx] ~ \exp \left[-\frac{1}{\hbar} S_{E}[\vec{x}] \right] \right|_{\vec{x}_{in},o}^{\vec{x}_{fi},T} ~ ,
\end{equation}
the classical action becomes the Euclidean classical action
\begin{equation}
\label{eq:EuclClassAct}
S_{E}[\vec{x}] = \int_{0}^{T} dt ~ \left\{ \frac{1}{2} m \dot{\vec{x}}^{2} + V(\vec{x}) \right\} ~ ,
\end{equation}
and the quantum action becomes the Euclidean quantum action
\begin{equation} 
\label{eq:EuclQuantAct}
\tilde{S}_{E}[\vec{x}] = \int_{0}^{T} dt ~ \left\{ \frac{1}{2} \tilde{m} \dot{\vec{x}}^{2} + \tilde{V}(\vec{x}) \right\} ~ .
\end{equation}
Our goal is to prove for $T \to \infty$ that the Euclidean transition amplitudes can be expressed in terms of the Euclidean quantum action, using a single trajectory,  
\begin{eqnarray}
\label{eq:DefEuclQuantAct}
&& G_{E}(\vec{x}_{fi},T; \vec{x}_{in},0) = \tilde{Z_{E}} 
\exp \left[ - \frac{1}{\hbar} 
\left. \tilde{\Sigma}_{E} \right|_{\vec{x}_{in},0}^{\vec{x}_{fi},T} \right] ~ ,
\nonumber \\
&& \left. \tilde{\Sigma}_{E} \right|_{\vec{x}_{in},0}^{\vec{x}_{fi},T} 
= \left. \tilde{S}_{E}[\tilde{\vec{x}}_{traj}] \right|_{\vec{x}_{in},0}^{\vec{x}_{fi},T}  
= \left. \int_{0}^{T} dt ~ \left\{ \frac{1}{2} \tilde{m} \dot{\tilde{\vec{x}}}_{traj}^{2} + \tilde{V}(\tilde{\vec{x}}_{traj}) \right\} \right|_{\vec{x}_{in}}^{\vec{x}_{fi}}  ~ .
\end{eqnarray}

\subsection{Necessary and sufficient conditions for existence of the quantum action}
For simplicity of notation we drop the subscript Euclidean. 
Let us make some assumptions on the potential $V(\vec{x})$:
Let $V(\vec{x}) \geq 0$. 
Let $V(\vec{x})$ be a smooth (sufficiently differentiable) function
of $\vec{x}$ and let $V(\vec{x}) \to \infty$ when $|\vec{x}| \to \infty$. Moreover, let $V(\vec{x})$ have a single minimum. Also we  assume that the ground state is non-degenerate. 
Because the Euclidean transition amplitude is given by a (Wiener) path integral with a positive Euclidean action, $G(\vec{y},T;\vec{x},0) \ge 0$ for all $\vec{x}$, $\vec{y}$. Hence we can introduce a real function $\eta$ such that 
\begin{equation}
\label{eq:DefEta}
G(\vec{y},T;\vec{x},0) = G_{0} \exp[-\eta(\vec{y},\vec{x})] ~ ,
\end{equation}
where $G_{0}$ is some constant (for fixed $T$) which takes care of the 
fact that $G$ has a dimension ($1/L^{D}$). 

Comparing the parametrisation of $G$ in terms of 
the function $\eta$, Eq(\ref{eq:DefEta}),
with its parametrisation in terms of the quantum action, Eq.(\ref{eq:DefEuclQuantAct}),
this looks similar. In order to prove it we need to establish that $\eta(\vec{y},\vec{x})$ can be expressed in terms of a local action $\tilde{S}$ evaluated along its trajectory $\tilde{\vec{x}}_{traj}$, such that
\begin{equation}
\label{eq:Identify}
\eta(\vec{y},\vec{x}) = \left. \frac{1}{\hbar} 
\tilde{S}[\tilde{\vec{x}}_{traj}] 
\right|_{\vec{x},t=0}^{\vec{y},t=T} ~~~ \mbox{and} ~~~ G_{0} = \tilde{Z} ~ . 
\end{equation}
This is a necessary and sufficient condition for the existence of the quantum action. In order to establish that those equations hold, we proceed in the following by establishing a number of equations equivalent to Eq.(\ref{eq:Identify}). At the end, we show that the last of those conditions can be satisfied. 

First, we compute partial derivatives of $\eta$ and $\Sigma$.
Consider the functional derivative up to first order of the action $\tilde{S}$ around the stationary point, admitting also variation of boundary pounts, 
denoted by
\begin{eqnarray}
&& \tilde{\vec{x}}(t) = \tilde{\vec{x}}_{traj}(t) + \tilde{\vec{h}}(t) ~ ,
\nonumber \\
&& \tilde{\vec{x}}_{traj}(t=0) = \vec{a}, ~  
\tilde{\vec{x}}_{traj}(t=T) = \vec{b} ~ ,
\nonumber \\
&& \tilde{\vec{h}}(t=0) = \delta \vec{a}, ~  
\tilde{\vec{h}}(t=T) = \delta \vec{b} ~ .
\end{eqnarray}
We obtain
\begin{equation}
\label{eq:VarAction}
\delta \tilde{S}[\tilde{\vec{x}}] 
= \tilde{\vec{p}}_{traj}(T) \cdot \delta \vec{b} - 
\tilde{\vec{p}}_{traj}(0) \cdot \delta \vec{a}  
+ O(\tilde{h}^{2}) ~ .
\end{equation}
On the other hand, one has
\begin{equation}
\label{eq:VarEta}
\delta \eta(\vec{b},\vec{a}) = \vec{\nabla}_{\vec{y}} ~ \eta(\vec{y},\vec{a})|_{\vec{y}=\vec{b}} 
\cdot \delta \vec{b} + 
\vec{\nabla}_{\vec{x}} ~ \eta(\vec{b},\vec{x})|_{\vec{x}=\vec{a}} \cdot \delta \vec{a} ~ .
\end{equation}
Comparing Eqs.(\ref{eq:VarAction},\ref{eq:VarEta}) for terms linear in $\delta \vec{a}$ and $\delta \vec{b}$, respectively, we find the following 
necessary and sufficient conditions for the existence of the quantum action,
\begin{equation}
\label{eq:Cond}
\left\{
\begin{array}{c} 
- \frac{1}{\hbar} \tilde{\vec{p}}_{traj}(0) =
\vec{\nabla}_{\vec{x}} ~ \frac{1}{\hbar} \tilde{\Sigma}|_{\vec{x}}^{\vec{y}} = \vec{\nabla}_{\vec{x}} ~ \eta(\vec{y},\vec{x})   
\\  
\frac{1}{\hbar} \tilde{\vec{p}}_{traj}(T) =
\vec{\nabla}_{\vec{y}} ~ \frac{1}{\hbar} \tilde{\Sigma}|_{\vec{x}}^{\vec{y}} = \vec{\nabla}_{\vec{y}} ~ \eta(\vec{y},\vec{x}) 
\end{array}
\right\}
~ \mbox{for all boundary points} ~ (\vec{y},\vec{x}) ~ .
\end{equation}
Eq.(\ref{eq:Cond}) is equivalent to
\begin{equation}
\label{eq:FunctIdent}
\frac{1}{\hbar} ~ \tilde{\Sigma}|_{\vec{x}}^{\vec{y}} = \eta(\vec{y},\vec{x}) ~~~ \mbox{modulo a global constant} ~ .
\end{equation}
The global constant can be absorbed into the constants $G_{0}$ and $\tilde{Z}$, respectively, and hence Eq.(\ref{eq:Cond}) is equivalent to Eq.(\ref{eq:Identify}).

\subsection{Use of energy conservation} 
It remains to be shown how that condition Eq.(\ref{eq:Cond}) can be satisfied. We do this by employing the principle of conservation of energy. The action 
given by Eq.(\ref{eq:EuclQuantAct})
describes a conservative system, i.e., the force is derived from a potential 
and energy is conserved. In imaginary time, energy conservation reads
\begin{equation}
- \tilde{T}_{kin} + \tilde{V} = \epsilon = \mbox{const} ~ ,
\end{equation}
which implies
\begin{equation}
\label{eq:EnerBalance}
\tilde{V}(\vec{b}) - \tilde{V}(\vec{a}) =
\frac{1}{2 \tilde{m}} (\tilde{\vec{p}}_{traj}^{fi})^{2}
- \frac{1}{2 \tilde{m}} (\tilde{\vec{p}}_{traj}^{in})^{2} ~ .
\end{equation}
Thus combining Eq.(\ref{eq:Cond}) and Eq.(\ref{eq:EnerBalance}), we find 
another necessary and sufficient condition for the existance of the quantum action: The quantum action exists and is local, if there is a mass $\tilde{m}$ and a local potential $\tilde{V}(\vec{x})$, such that 
\begin{equation}
\label{eq:Cond2}
\frac{2 \tilde{m}}{\hbar^{2}} \left[ \tilde{V}(\vec{b}) - \tilde{V}(\vec{a}) \right] 
= \left( \vec{\nabla}_{\vec{y}} ~ \eta(\vec{y},\vec{a})|_{\vec{y}=\vec{b}} \right)^{2} 
- \left( \vec{\nabla}_{\vec{x}} ~ \eta(\vec{b},\vec{x})|_{\vec{x}=\vec{a}} \right)^{2} ~~~ \mbox{holds for all} ~ \vec{a}, \vec{b} ~ .
\end{equation}

\subsection{Feynman-Kac limit}
Finally, we want to show that Eq.(\ref{eq:Cond2}) can be satisfied in the limit
$T \to \infty$. In this limit holds the Feynman-Kac formula 
\begin{equation} 
G(\vec{y},T;\vec{x},0) \longrightarrow_{T \to \infty} \langle \vec{y} | \psi_{gr} \rangle 
e^{-E_{gr}T/\hbar} \langle \psi_{gr} | \vec{x} \rangle ~ ,
\end{equation}
where $\psi_{gr}$ is the ground state wave function and $E_{gr}$ the ground state energy. Here we use the assumption that the ground state is not degenerate. 
Eq.(\ref{eq:DefEta}) implies
\begin{equation}
G_{0} e^{-\eta(\vec{y},\vec{x})} \longrightarrow_{T \to \infty} \langle \vec{y} | \psi_{gr} \rangle 
e^{-E_{gr}T/\hbar} \langle \psi_{gr} | \vec{x} \rangle ~ .
\end{equation}
From this we compute
\begin{eqnarray}
&&\vec{\nabla}_{\vec{y}} ~ \eta(\vec{y},\vec{x})|_{\vec{y}=\vec{b},\vec{x}=\vec{a}} 
\longrightarrow_{T \to \infty} 
= - \frac{ \vec{\nabla} \psi_{gr}(\vec{b})}{\psi_{gr}(\vec{b})} ~ .
\nonumber \\
&&\vec{\nabla}_{\vec{x}} ~ \eta(\vec{y},\vec{x})_{\vec{y}=\vec{b},\vec{x}=\vec{a}} 
\longrightarrow_{T \to \infty} 
- \frac{\vec{\nabla} \psi_{gr}(\vec{a})}{\psi_{gr}(\vec{a})} ~ .
\end{eqnarray}
Then the general condition, Eq.(\ref{eq:Cond2}) becomes
\begin{equation} 
\label{eq:QuantPot}
\frac{2 \tilde{m}}{\hbar^{2}} [ \tilde{V}(\vec{b}) - \tilde{V}(\vec{a}) ] 
\longrightarrow_{T \to \infty} \left( \frac{\vec{\nabla} \psi_{gr}(\vec{b})}{\psi_{gr}(\vec{b})} \right)^{2} 
- \left( \frac{\vec{\nabla} \psi_{gr}(\vec{a})}{\psi_{gr}(\vec{a})} \right)^{2} ~ \mbox{for all} ~ 
\vec{a}, \vec{b} ~ .
\end{equation}
This holds if we can find $\tilde{m}$ and $\tilde{V}(\vec{x})$, such
that
\begin{equation} 
\label{eq:FinalCond}
\frac{2 \tilde{m}}{\hbar^{2}} \left( \tilde{V}(\vec{x}) - \tilde{V}_{0} \right) 
= \left( \frac{\vec{\nabla} \psi_{gr}(\vec{x})}{\psi_{gr}(\vec{x})} \right)^{2} ~ 
\mbox{for all} ~ \vec{x} ~ ,
\end{equation}
where $\tilde{V}_{0}$ denotes the minimum of the potential. This condition can be satisfied. This establishes the existence of a local quantum action and finishes the proof.

\subsection{Relation between classical and quantum potential}
We continue to work in the limit $T \to \infty$. 
Let us consider the stationary Schr\"odinger equation for the ground state, which can be written in the form
\begin{equation}
\label{eq:SchrodEqGround}
\frac{ \Delta \psi_{gr}(\vec{x}) }{ \psi_{gr}(\vec{x}) } =
\frac{2m}{\hbar^{2}} [ V(\vec{x}) - E_{gr} ] ~ ,
\end{equation}
where $m$ is the classical mass and $V$ is the classical potential. Because 
the transition amplitude is positive, also the ground state wave function obeys
$\psi_{gr}(\vec{x}) \ge 0$ for all $\vec{x}$. 
Hence we can define the function $U(\vec{x})$ 
\begin{equation}
U(\vec{x}) = \log \psi_{gr}(\vec{x}) ~ .
\end{equation}
Using Eqs.(\ref{eq:FinalCond},\ref{eq:SchrodEqGround}) we compute
\begin{equation}
\Delta U(\vec{x}) = 
\frac{2 \tilde{m}}{\hbar^{2}} [ \tilde{V}(\vec{x}) - \tilde{V}_{0} ]
+ \frac{2m}{\hbar^{2}} [ V(\vec{x}) - E_{gr} ] ~ .
\end{equation}
It can be cast into the form
\begin{eqnarray}
\label{eq:Riccati1}
&& \Delta U(\vec{x}) + (\vec{\nabla} U(\vec{x}))^{2} = 
\frac{2m}{\hbar^{2}} [ V(\vec{x}) - E_{gr} ] ~ ,
\nonumber \\
&& \frac{2 \tilde{m}}{\hbar^{2}} [ \tilde{V}(\vec{x}) - \tilde{V}_{0} ] = (\vec{\nabla} U)^{2} ~ .
\end{eqnarray} 
The first equation is a generalized Riccati-type differential equation for the function $U(\vec{x})$. 
From its solution one can obtain the the quantum potential, more precisely $\tilde{m} \tilde{V}(\vec{x})$. Hence, Eqs.(\ref{eq:Riccati1}) constitute a transformation rule 
\begin{equation}
m V(\vec{x}) \to \tilde{m} \tilde{V}(\vec{x}) ~ .
\end{equation}
From the point of view of numerical solution, it may be advantageous to cast Eq.(\ref{eq:Riccati1}) in a slightly different form. Let us define
\begin{equation}
\label{eq:DefW}
W(\vec{x}) = \vec{\nabla} U(\vec{x}) ~ .
\end{equation}
Then the generalized Riccati differential equation becomes
\begin{eqnarray}
\label{eq:Riccati2}
&& \vec{\nabla} \cdot \vec{W}(\vec{x}) + (\vec{W}(\vec{x}))^{2} = 
\frac{2m}{\hbar^{2}} [ V(\vec{x}) - E_{gr} ]
\nonumber \\
&& \frac{2 \tilde{m}}{\hbar^{2}} [ \tilde{V}(\vec{x}) - \tilde{V}_{0} ] = 
(\vec{W}(\vec{x}))^{2} ~ .
\end{eqnarray} 

\subsection{Relation between quantum action and potentials in supersymmetric quantum mechanics}

Starting with Witten's work \cite{Witten81} it was recognized that supersymmetry can be applied to quantum mechanics. In Ref.\cite{Bernstein} the correspondence between the 1-D Fokker-Planck equation with a bi-stable potential and supersymmetric quantum mechanics has been shown to facilitate the calculation of a small eigenvalue that controls the rate at which equilibrium is approached.
Kumar et al. \cite{Kumar} used the supersymmetric approach to compute the tunneling exchange integral $t=E_{1}-E_{0}$ and pointed out the Riccati differential equation between the classical and supersymmetric partner potential. Let us consider a 1-D Hamiltonian system, which in the notation of Ref.\cite{Keung} and using $\hbar=2m=1$ is given by
\begin{equation}
H_{-} = - \frac{d^{2}}{dx^{2}} + V_{-}(x) ~ ,
\end{equation}
where $V_{-}$ denotes the standard classical potential.
One can construct a supersymmetric partner potential $V_{+}(x)$,
\begin{equation}
V_{+}(x) = V_{-}(x) - 2 \frac{d}{dx} \frac{\psi'_{gr}}{\psi_{gr}} 
= - V_{-}(x) + 2 (\frac{\psi'_{gr}}{\psi_{gr}})^{2} ~ .
\end{equation}
Defining a superpotential $W_{s}(x)$ by
\begin{equation}
\label{eq:DefWSuper}
W_{s}(x) = - \frac{\psi'_{gr}}{\psi_{gr}} ~ ,
\end{equation}
one obtains 
\begin{equation}
\label{eq:RiccatiSupSym}
V_{\pm}(x) = W_{s}^{2}(x) \pm \frac{dW_{s}}{dx} ~ ,
\end{equation}
which is a Riccati-type equation. The following similarities are apparent. Comparing Eqs.(\ref{eq:DefW},\ref{eq:DefWSuper}) one finds
\begin{equation}
W(x) = - W_{s}(x) ~ ,
\end{equation}
which satisfy the corresponding Riccati differential equations (\ref{eq:Riccati2},\ref{eq:RiccatiSupSym}). 
The following remarks are in order. 
In the super-symmetric approach the mass is identical to the classical mass, while in the quantum action, the mass gets tuned, in general. In contrast to the super-symmetric quantum Hamiltonian, the quantum action gives a parametrisation of quantum transition amplitudes (see Eq.(\ref{eq:DefQAction})), which requires a tuning of the mass.

\section{Model and its quantum action}
\label{sec:Model}
\begin{figure*}[ht]
\begin{center}
\includegraphics[scale=0.50,angle=180]{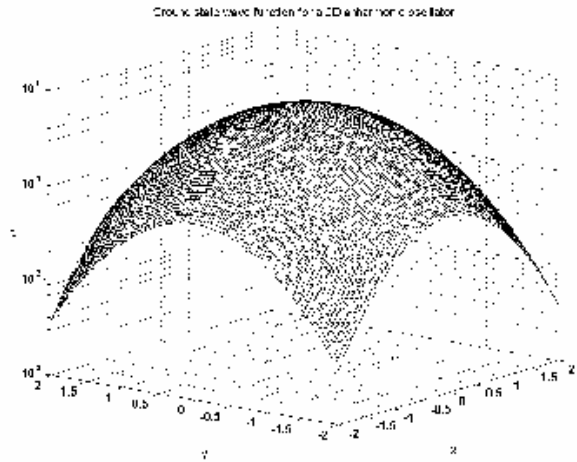}
\end{center}
\caption{Ground state wave function $\psi_{gr}(x,y)$.}
\label{fig:GroundState}
\end{figure*}
\begin{figure*}[ht]
\begin{center}
\includegraphics[scale=0.50,angle=180]{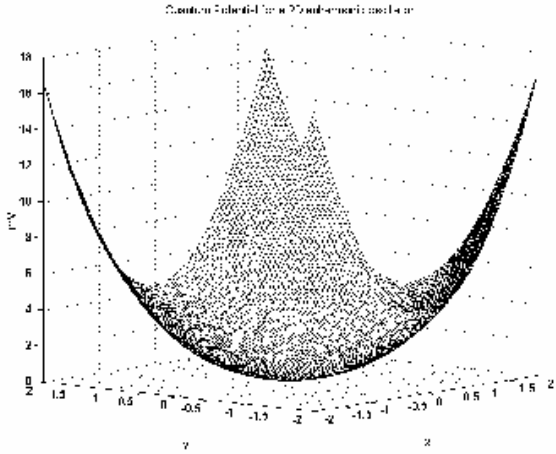}
\end{center}
\caption{Quantum potential $\tilde{m}(\tilde{V}(x,y) - \tilde{V}_{min})$.}
\label{fig:QuantPot}
\end{figure*}
\begin{table}[ht]
\caption{Parameters of classical action vs. quantum action. $v_{22}=0.25$, $T=4.5$.}
\label{tab:ParamAction}
\begin{center}
\begin{tabular}{||c|c|c||}
\hline
Parameter & Class. Action & Quant. Action \\ \hline
$ m $ & 1 & $0.976(8)$ \\ \hline
$ v_{0} $ & 0 & $1.3992(4)$ \\ \hline
$ v_{2} $ & 0.5 & $0.5684(3)$ \\ \hline
$ v_{22} $ & 0.25 & $0.2469(4)$ \\ \hline
$ v_{4} $ & 0 & $-0.00067(7)$ \\ \hline
\end{tabular}
\end{center}
\end{table}
In order to study full chaos the so-called K-system (2-D Hamiltonian, 
potential $V=x^{2}y^{2}$) is widely used.
It is almost globally chaotic, having small islands of stability \cite{Dahlquist}. In order to study mixed dynamics (entangling chaos and regular islands) it is numerically convenient to consider the following classical 
system \cite{Pullen}, 
\begin{equation}
S = \int_{0}^{T} dt ~ \frac{1}{2} m (\dot{x}^{2} + \dot{y}^{2}) - V(x,y) ~ , ~~~ V = v_{2}(x^{2} + y^{2}) + v_{22} x^{2}y^{2} ~ .
\end{equation}
The parameters of the classical action are given in Tab.[\ref{tab:ParamAction}]. The ground state wave function of the quantum system, obtained by solving the Schr\"odinger equation, is depicted in Fig.[\ref{fig:GroundState}]. 
The quantum action (see Eq.(\ref{eq:DefQAction})) takes the form
\begin{equation}
\tilde{S} = \int_{0}^{T} dt ~ \frac{1}{2} \tilde{m} (\dot{x}^{2} + \dot{y}^{2}) - \tilde{V}(x,y) ~ .
\end{equation}
We need to determine the quantum mass (tuned mass) $\tilde{m}$ and the quantum potential $\tilde{V}(x,y)$. For large imaginary time $T$ we have analytical results on the form of $\tilde{V}$. The function $\tilde{m}(\tilde{V}(x,y)-\tilde{V}_{min})$ is given by Eq.(\ref{eq:FinalCond}) or equivalently by the solution of Eq.(\ref{eq:Riccati2}). This is depicted in Fig.[\ref{fig:QuantPot}]. 
We need to determine $\tilde{m}$ and $\tilde{V}$ separately and also the parameter $\tilde{Z}$ of the quantum action. Those have been computed numerically in a non-perturbative way from a global fit to transition amplitudes in imaginary time. 
For this purpose we made a polynomial ansatz for the quantum action of the following form,
\begin{eqnarray}
\tilde{V}(x,y) &=& \tilde{v}_{0} + \tilde{v}_{11} xy + \tilde{v}_{2} (x^{2}+y^{2}) + \tilde{v}_{22} x^{2}y^{2} + \tilde{v}_{13} (xy^{3} + x^{3}y)
+ \tilde{v}_{4} (x^{4} + y^{4}) 
\nonumber \\
&+& \tilde{v}_{24} (x^{2}y^{4} + x^{4}y^{2})
+ \tilde{v}_{44} x^{4}y^{4} ~ .
\end{eqnarray}
We have computed 
transition amplitudes $G(\vec{x}_{fi},t_{fi}; \vec{x}_{in},t_{in})$ by solving the Schr\"{o}dinger equation in imaginary time keeping the transition time $T$ fixed. The determination of unknown parameters of the quantum action, like   $\tilde{Z}$, $\tilde{m}$, etc.,
requires as many equations as unknowns. In practice we use much more equations (overdetermined system), by considering a large set of pairs of initial and final boundary points $(\vec{x}_{j},\vec{x}_{i})$ and the corresponding transition amplitudes. 
Next we make an initial guess for the unknown parameters, say
$\tilde{Z}^{(0)}$, $\tilde{m}^{(0)}$, etc. and solve the Euler-Lagrange equations of motion derived from the quantum action $\tilde{S}^{(0)}$, for each pair of boundary points. Then we compute the error 
\begin{equation}
\epsilon = \sum_{i,j=1}^{N} \left| G(\vec{x}_{i},T;\vec{x}_{j},0) - 
\tilde{Z}^{(0)} \exp \left[ \frac{i}{\hbar} \left. \tilde{S}^{(0)}[\tilde{\vec{x}}^{(0)}_{traj}] 
\right|_{\vec{x}_{j},0}^{\vec{x}_{i},T} \right] \right|^{2} ~ .
\end{equation}
In order to find eventually the physical values of the unknown parameters we apply a variational principle, i.e. we search in the multi-parameter space to minimize the error globally (over all pairs of boundary points). For more details see Ref.\cite{Q1}.
\begin{figure*}[htp]
\vspace{2pt}
\begin{center}
\includegraphics[scale=0.30,angle=-90]{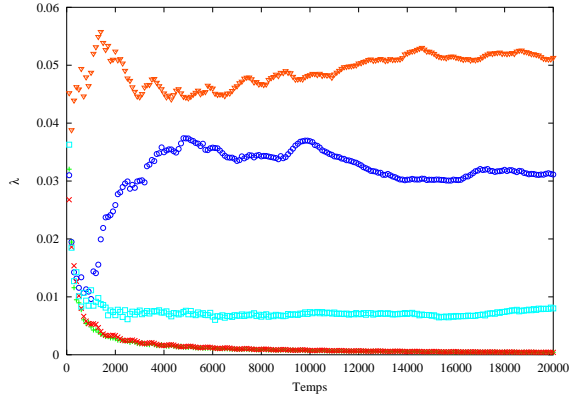}
\end{center}
\caption{Typical time evolution of Lyapunov exponents. $v_{22}=0.25$. $E=2$.}
\label{fig:L25_TimeEvol}
\end{figure*}
As a results, the terms $\tilde{v}_{11}$, $\tilde{v}_{13}$, $\tilde{v}_{4}$, $\tilde{v}_{24}$, $\tilde{v}_{44}$, were found to be quite small or compatible with zero.
The parameters of the corresponding quantum action for transition time $T=4.5$ (large compared to dynamical time scale $T_{sc}=1/E_{gr}$) and parameter $v_{22}=0.25$ (which controls chaos) are shown in Tab.[\ref{tab:ParamAction}]. 
The bracket gives the estimated error of the fit. 
The numerical results presented below refer to $v_{22}=0.25$ and $v_{22}=0.05$. 
\begin{figure*}[htp]
\vspace{2pt}
\begin{center}
\includegraphics[scale=0.45,angle=-270]{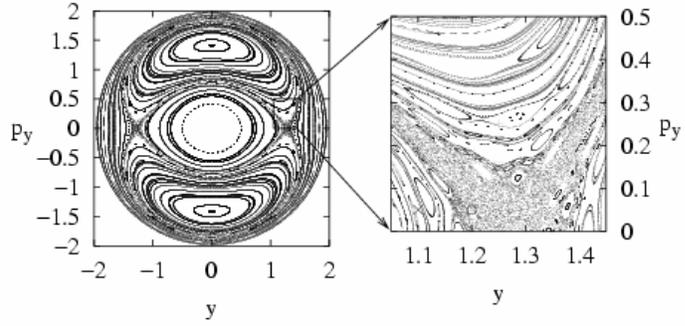}
\vspace{-5pt}
\includegraphics[scale=0.45,angle=-270]{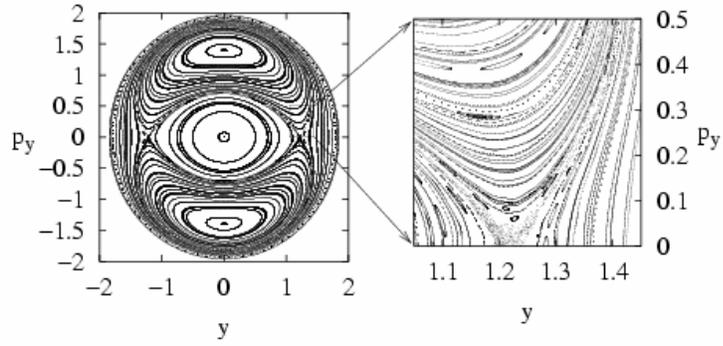}
\end{center}
\caption{Poincare section of classical system (top) and quantum system (bottom). $v_{22}=0.25$. $E=2$.}
\label{fig:Poin_E2_L25}
\end{figure*}
\begin{figure*}[htp]
\vspace{2pt}
\begin{center}
\includegraphics[scale=0.35,angle=180]{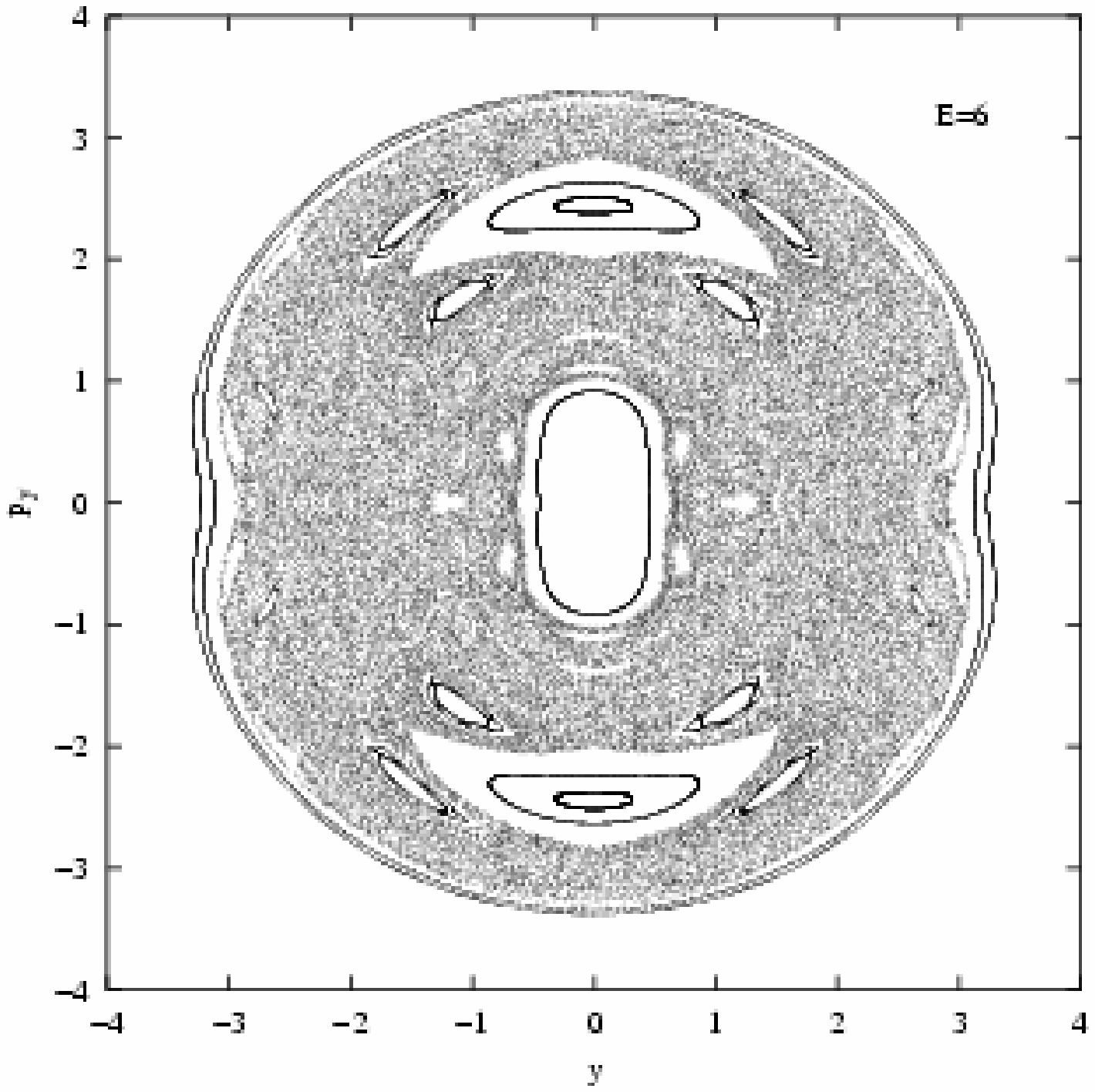}
\vspace{-5pt}
\includegraphics[scale=0.35,angle=180]{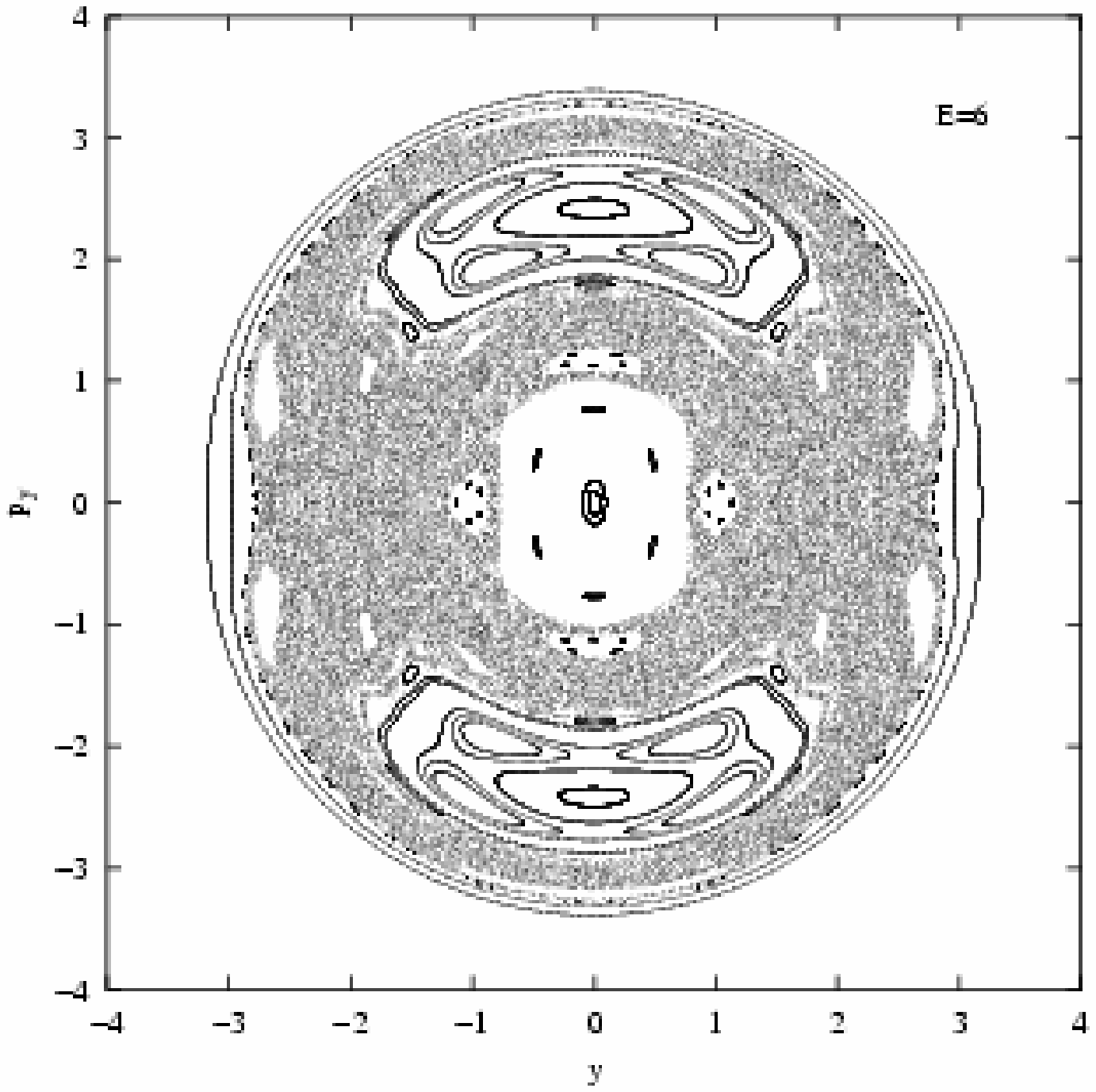}
\end{center}
\caption{Poincare section of classical system (left) and quantum system (right). $v_{22}=0.25$. $E=6$.}
\label{fig:Poin_E6_L25}
\end{figure*}
Before we present our comparison of classical to quantum chaos, we ask what behavior do we expect? In the 1-D quartic potential quantum effects produce a strong positive quadratic term in the quantum action \cite{Q1}.
In the 1-D double well potential, the quantum potential becomes softened, 
i.e., it has a lower barrier and closer minima. As a consequence, the instanton solutions of the quantum action are softer than the corresponding classical instantons \cite{Q2}. Are such softening effects also occuring in chaotic phenomena?  
\begin{figure}[htp]
\vspace{9pt}
\begin{center}
\includegraphics[scale=0.35,angle=180]{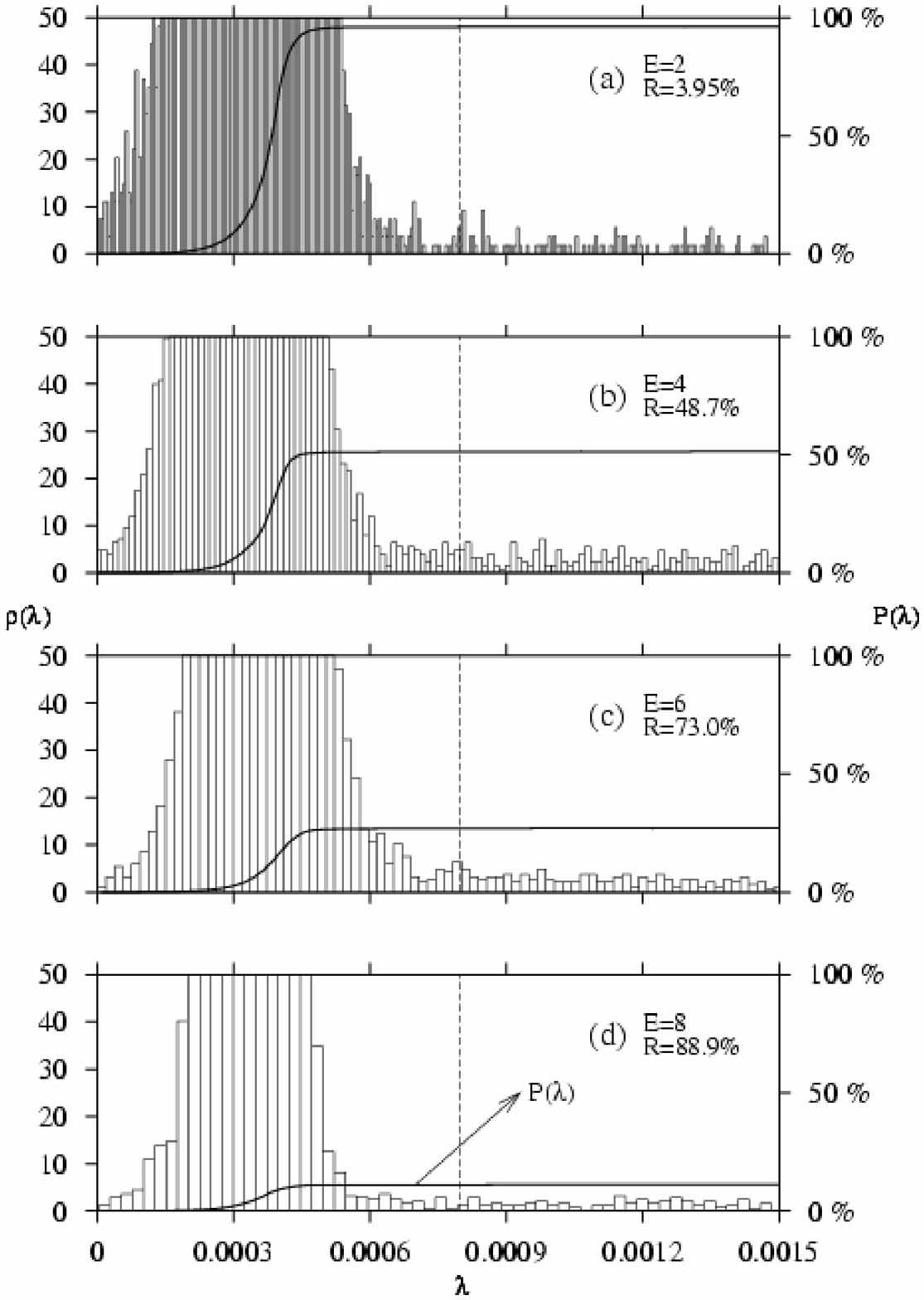}
\includegraphics[scale=0.35,angle=180]{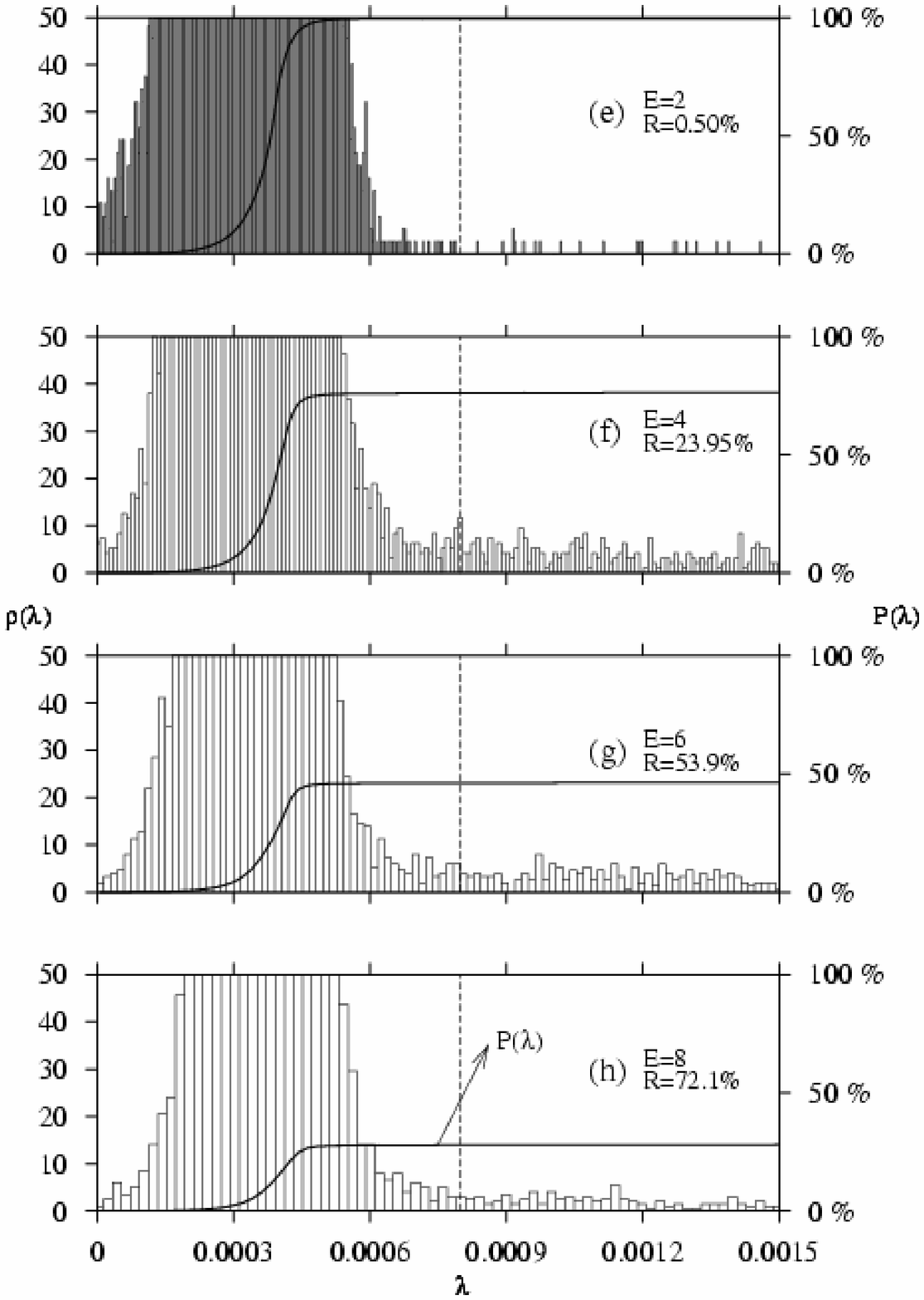}
\end{center}
\caption{Distribution of Lyapunov exponents near $\lambda=0$ for different energies $E$. Classical system (left column) and quantum system (right column). $v_{22}=0.25$.}
\label{fig:Zero_Lyap_4xE_L25}
\end{figure}
\begin{figure}[htp]
\vspace{9pt}
\begin{center}
\includegraphics[scale=0.35,angle=180]{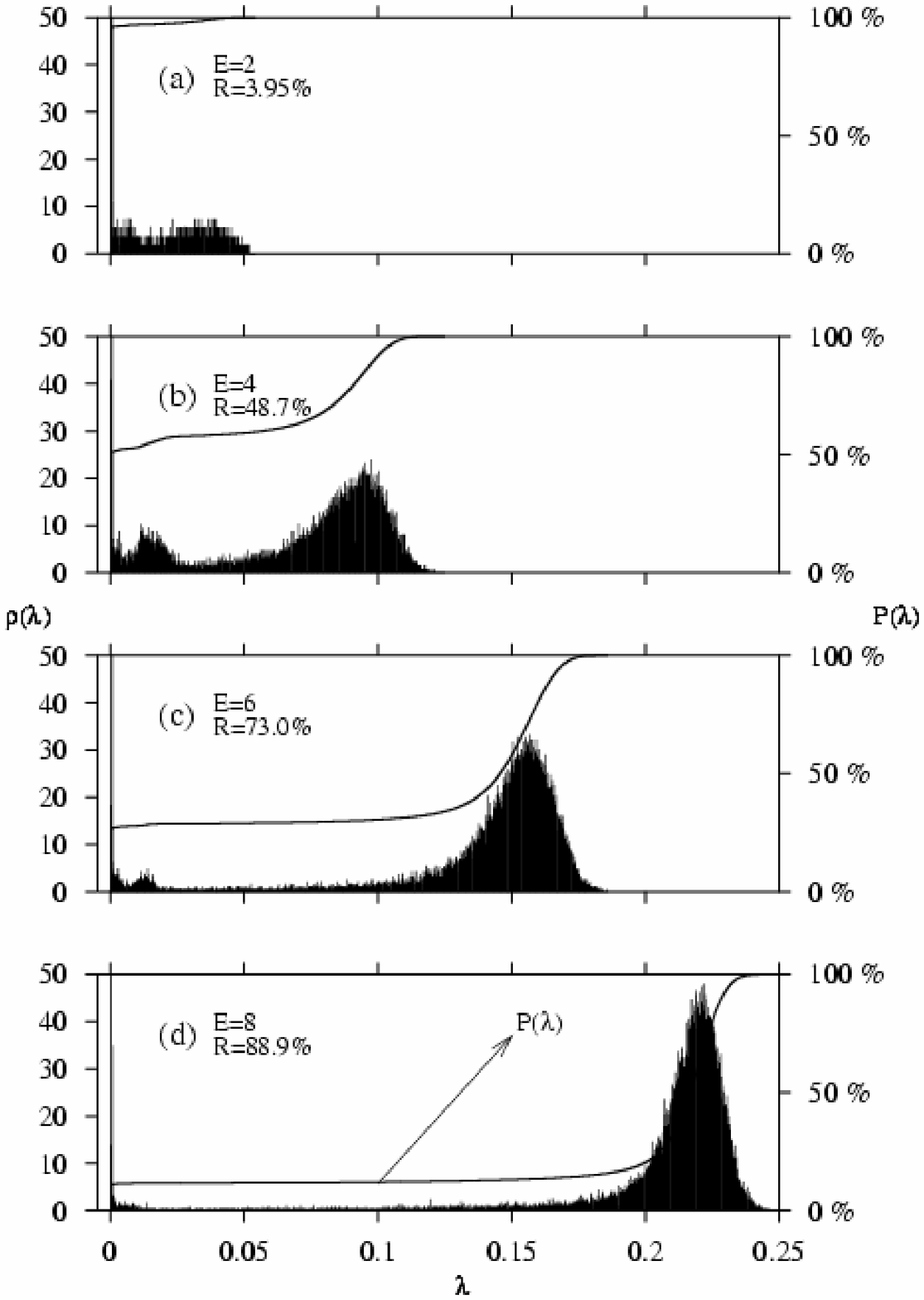}
\includegraphics[scale=0.35,angle=180]{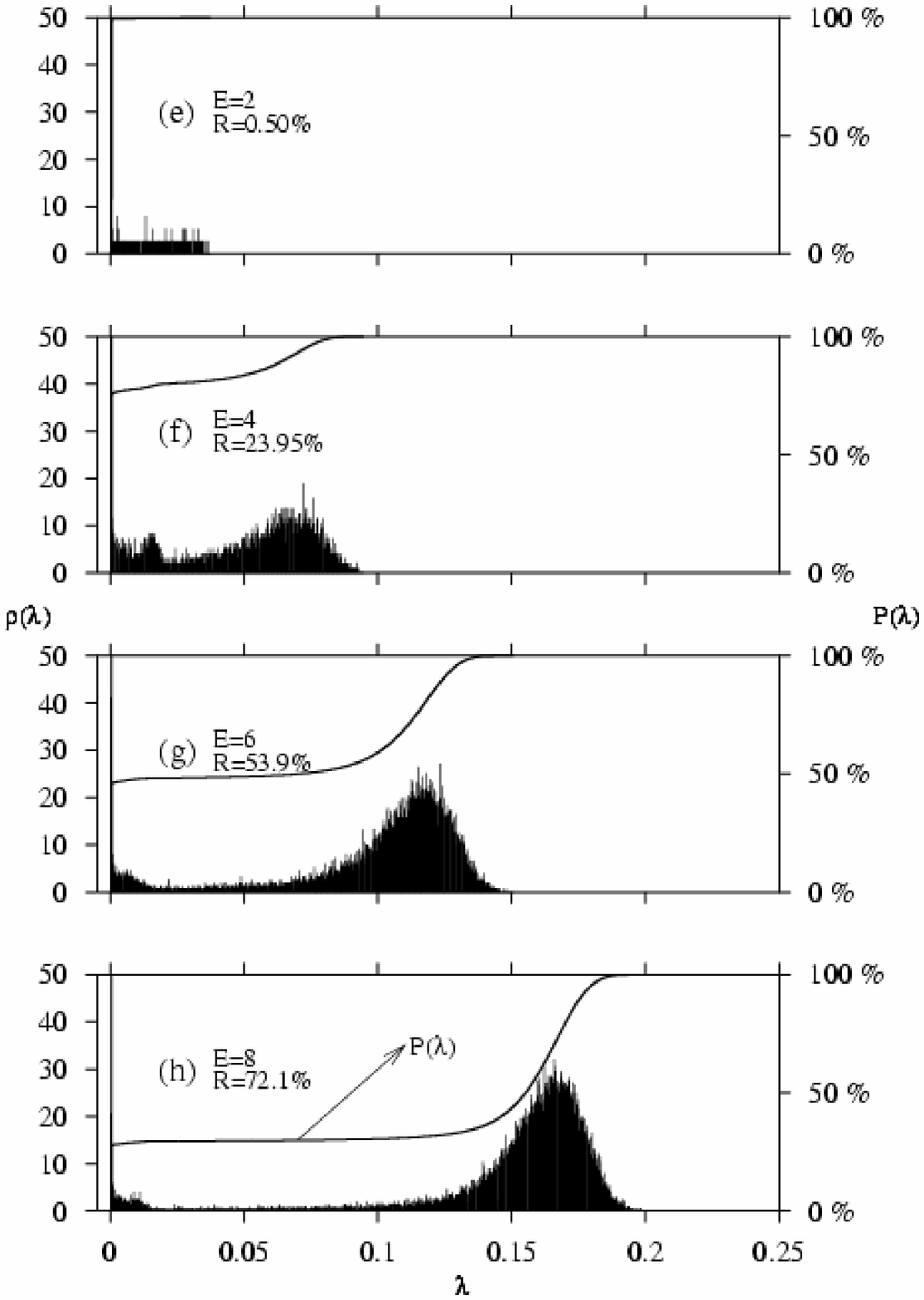}
\end{center}
\caption{Distribution of positive Lyapunov exponents. Classical system (left column) and quantum system (right column). $v_{22}=0.25$.}
\label{fig:Pos_Lyap_4xE_L25}
\end{figure}

\section{Classical chaos vs. quantum chaos}
\label{sec:CompareChaos}
The notion of chaotic versus regular phase space is based on Lyapunov exponents 
($\lambda > 0$ versus $\lambda \le 0$). Numerically it is difficult to distinguish a small positive from a zero Lyapunov exponent. In order to discriminate those cases we followed the time evolution for a long time. 
We used $T_{c}=20000$ to measure $\lambda$. This is depicted in Fig.[\ref{fig:L25_TimeEvol}].
A comparison of Poincar\'e sections of the classical and the quantum system is shown in Fig.[\ref{fig:Poin_E2_L25}] for $v_{22}=0.25$ and $E=2$. 
One observes a smaller number of chaotic trajectories in the quantum case than in the classical case. In particular, this is visible near to the hyperbolic fixed points (see insert). The same comparison for some higher energy ($E=6$) is shown in Fig.[\ref{fig:Poin_E6_L25}]. With increasing energy the difference between classical and quantum phase space becomes more pronounced.
In order to get a quantitative measure for chaotic versus regular phase space, 
we have chosen randomly a large number of initial conditions and computed the finite time Lyapunov exponent for each such trajectory.
The distribution of these Lyapunov exponents in the neighborhood of $\lambda=0$ 
is shown in Fig.[\ref{fig:Zero_Lyap_4xE_L25}], and those for 
$0 < \lambda < 0.25$ is shown in Fig.[\ref{fig:Pos_Lyap_4xE_L25}].
Fig.[\ref{fig:Zero_Lyap_4xE_L25}] shows the distribution of Lyapunov exponents for different energies for the classical and quantum system. While $\lambda=0$ corresponds to regular behavior, $\lambda >0$ indicates chaotic behavior. For the purpose to distinguish numerically the two regimes, we used this distribution to define some cut-off $\lambda_{c}$ (vertical dashed line in Fig.[\ref{fig:Zero_Lyap_4xE_L25}]). Fig.[\ref{fig:Zero_Lyap_4xE_L25}] also shows the cumulative distribution $P(\lambda)$ (full line), which is a measure for the degree of regularity. As can be seen this curve is higher for the quantum systen compared to the classical system. 
Moreover, Fig.[\ref{fig:Zero_Lyap_4xE_L25}] displays the coefficient $R$ which denotes the ratio of chaotic phase space over total phase space.
\begin{figure}[htp]
\vspace{9pt}
\begin{center}
\includegraphics[scale=0.35,angle=180]{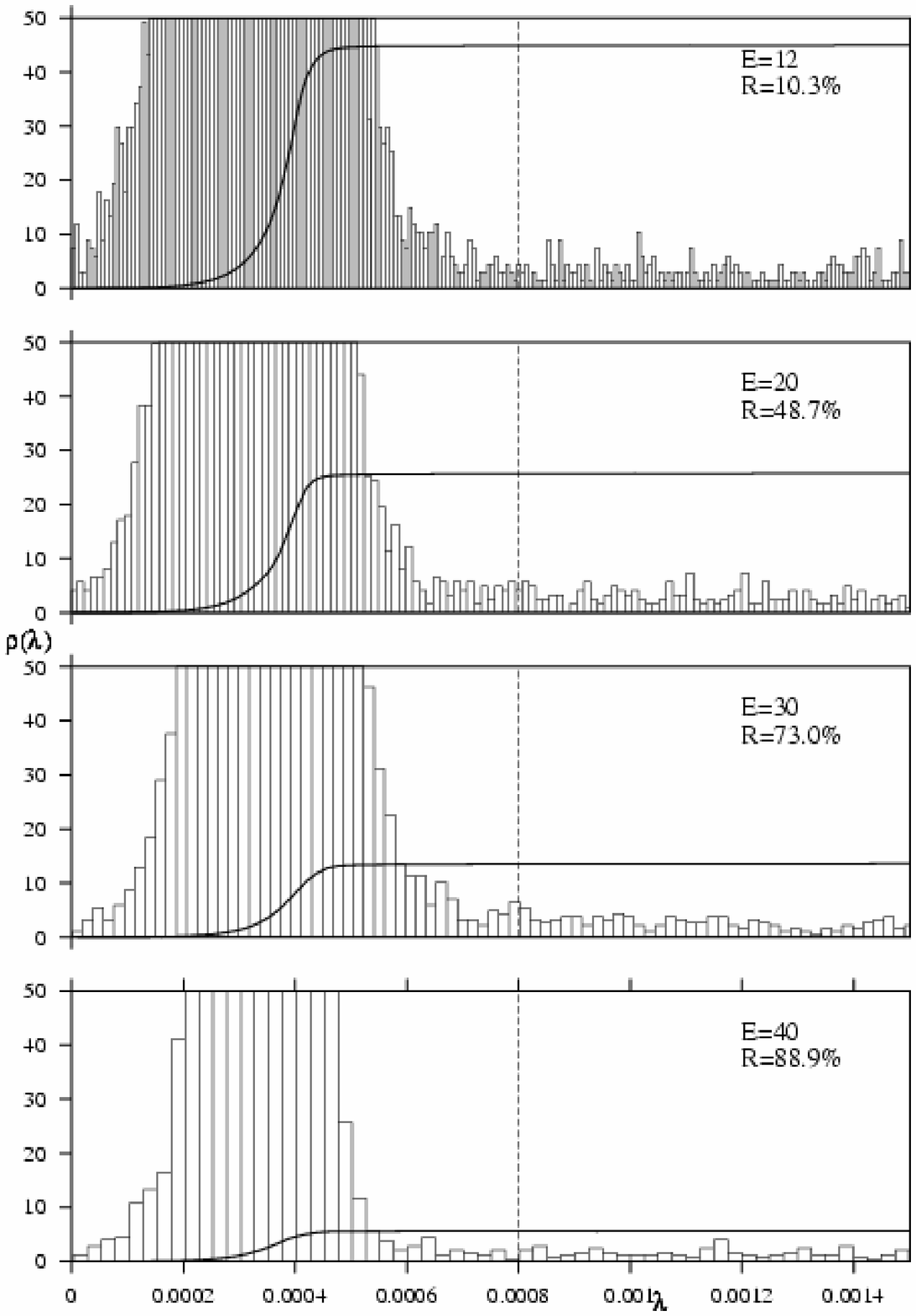}
\includegraphics[scale=0.35,angle=180]{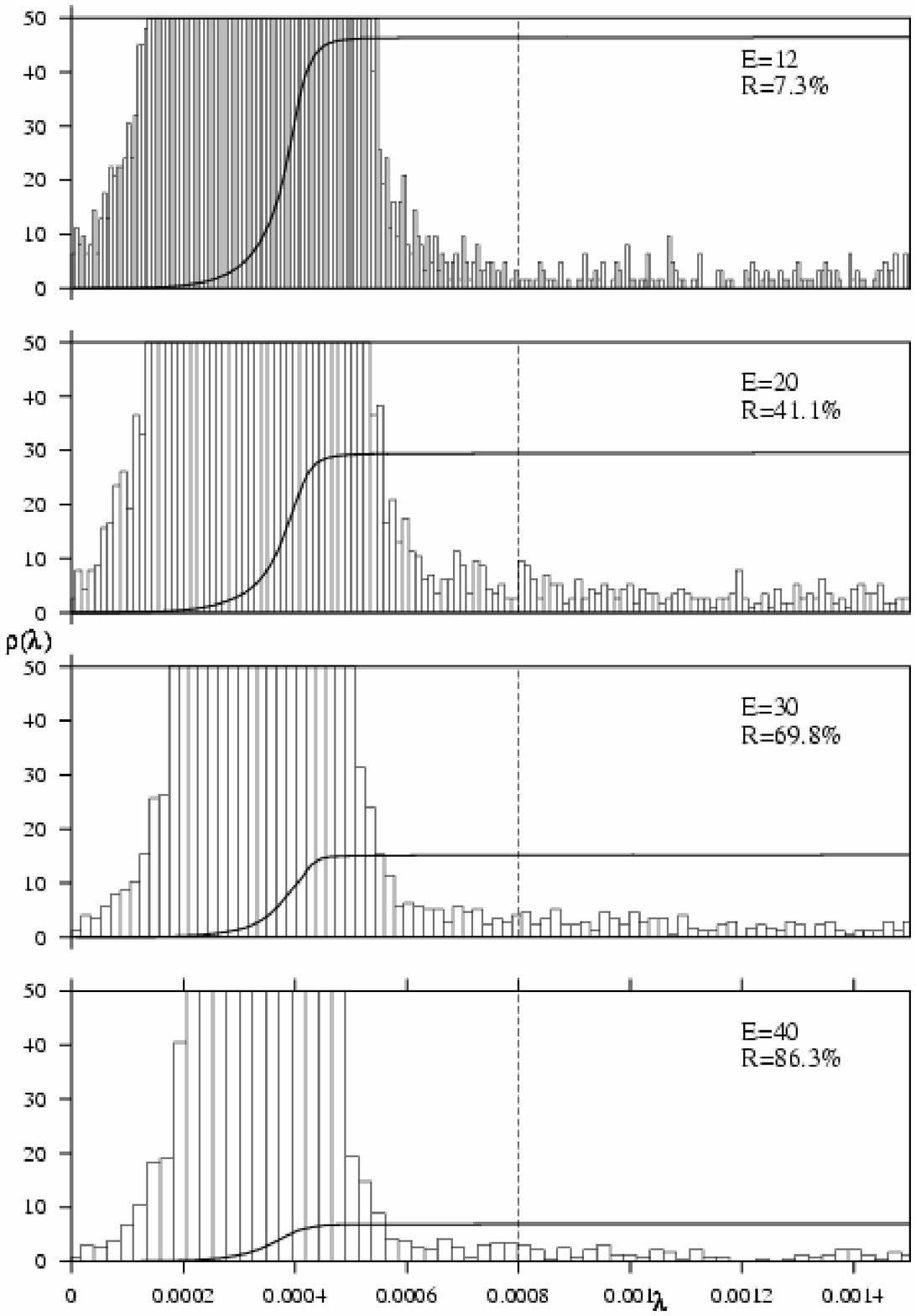}
\end{center}
\caption{Distribution of Lyapunov exponents near $\lambda=0$ for different energies $E$. Classical system (left column) and quantum system (right column). $v_{22}=0.05$.}
\label{fig:Zero_Lyap_4xE_L05}
\end{figure}
\begin{figure}[htp]
\vspace{9pt}
\begin{center}
\includegraphics[scale=0.35,angle=180]{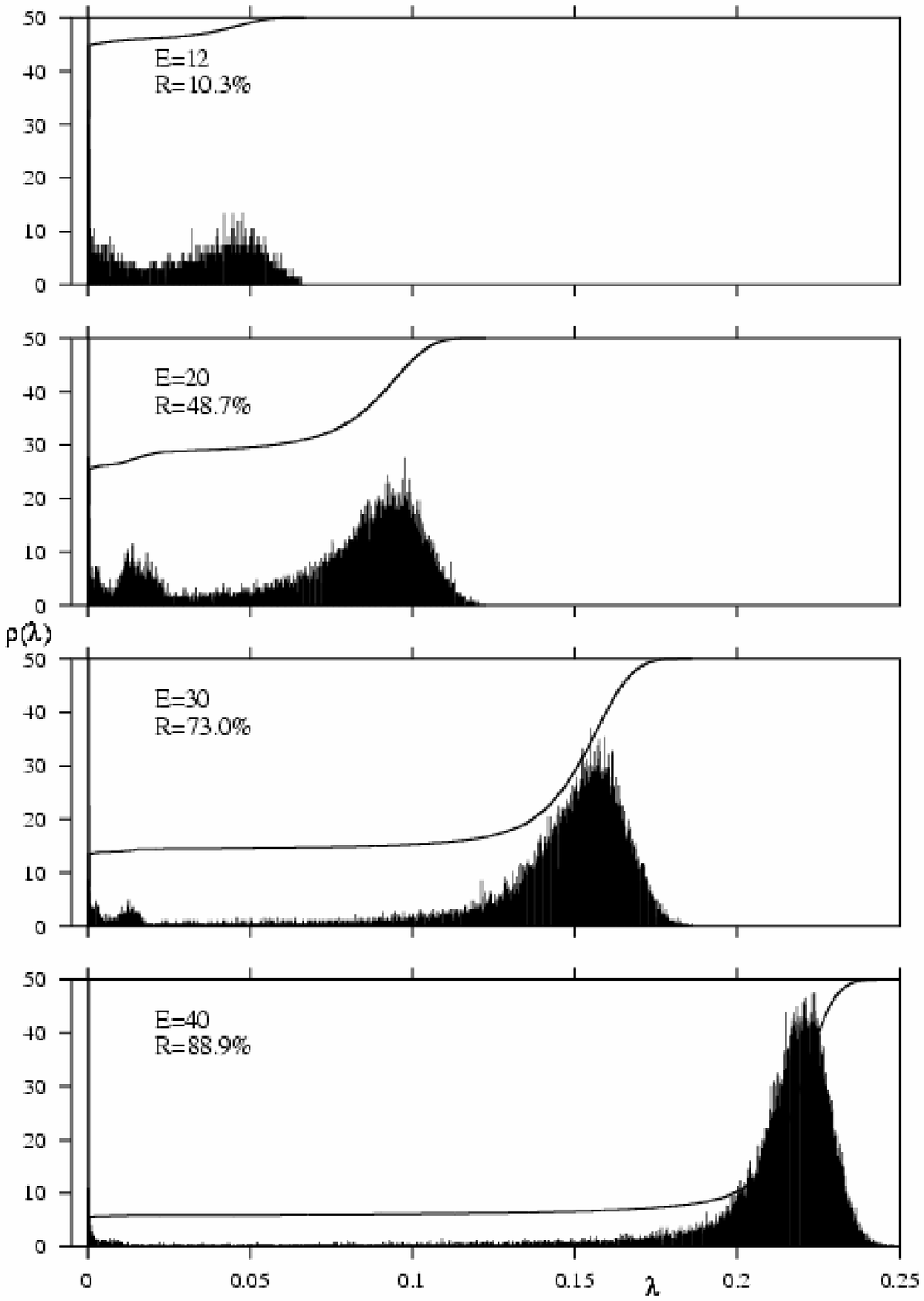}
\includegraphics[scale=0.35,angle=180]{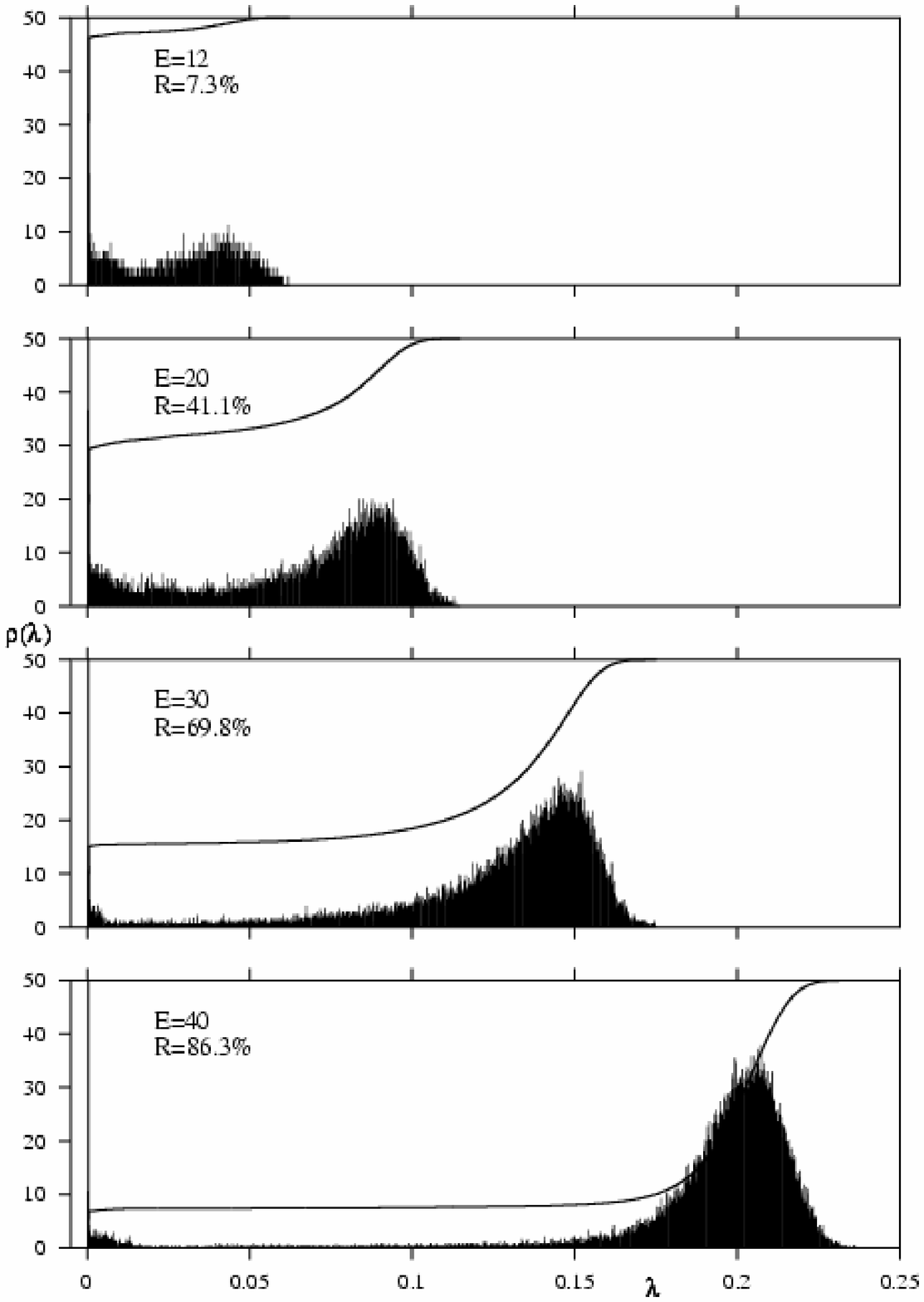}
\end{center}
\caption{Distribution of positive Lyapunov exponents. Classical system (left column) and quantum system (right column). $v_{22}=0.05$.}
\label{fig:Pos_Lyap_4xE_L05}
\end{figure}
The distribution of positive Lyapunov exponents is displayed in Fig.[\ref{fig:Pos_Lyap_4xE_L25}].
First, one observes a pronounced peak near zero, which has been magnified in 
Fig.[\ref{fig:Zero_Lyap_4xE_L25}]. Second, with increase of energy, the system becomes more chaotic, i.e. $R$ increases. Third, with increasing energy, the distribution of positive Lyapunov exponents seems to approach a Gaussian shape.

\begin{figure}[htp]
\vspace{9pt}
\begin{center}
\includegraphics[scale=0.3,angle=0]{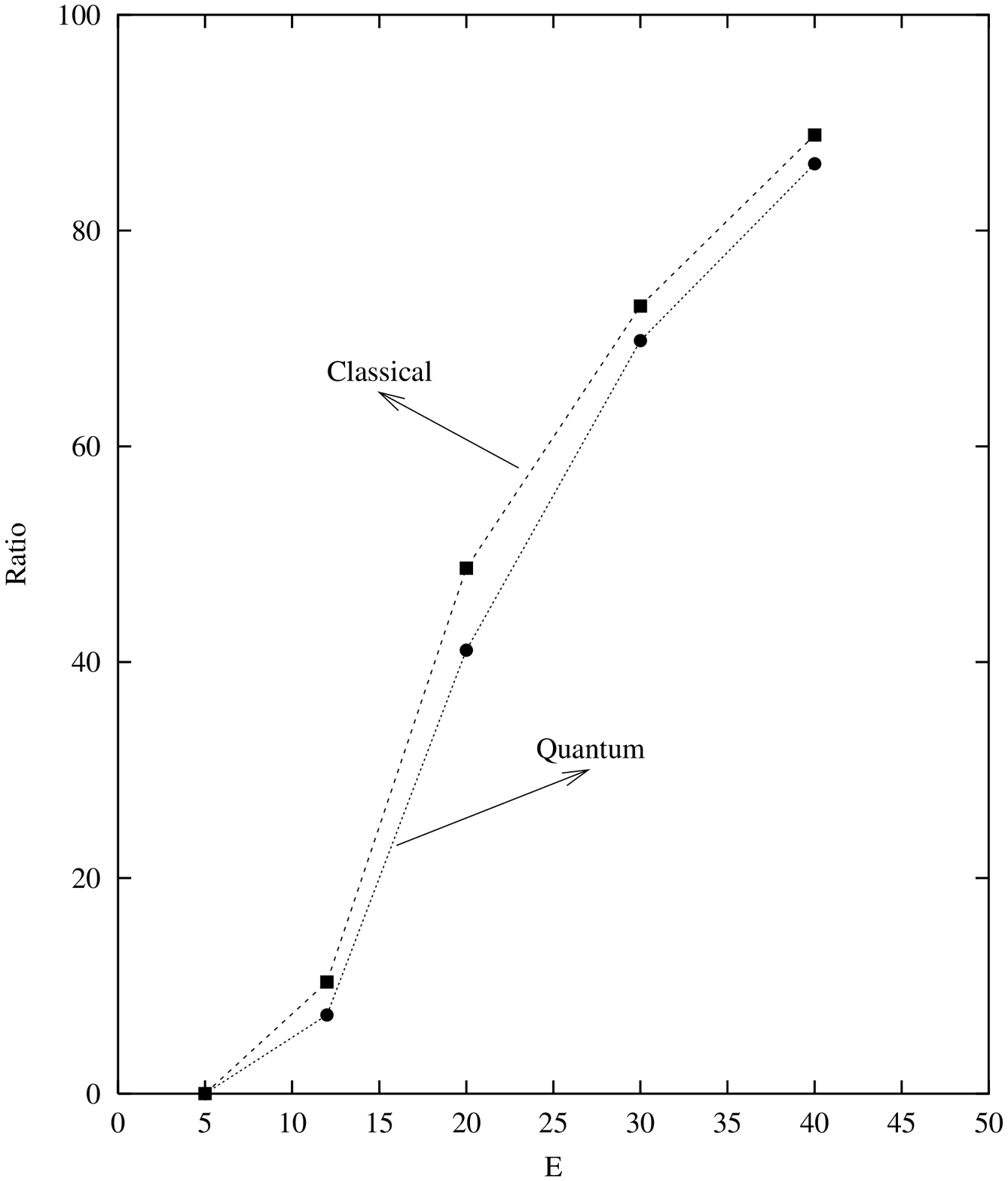}
\includegraphics[scale=0.3,angle=0]{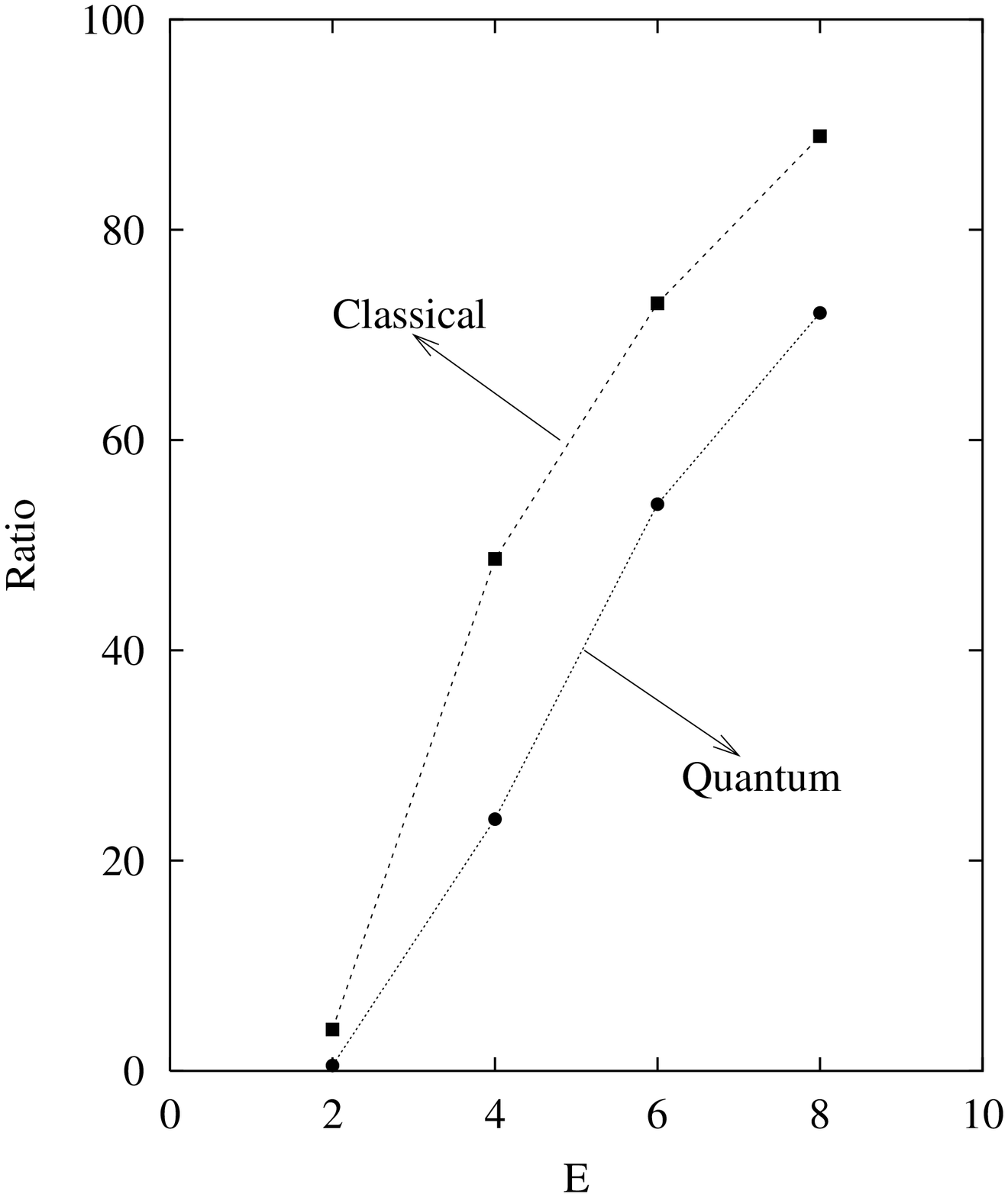}
\end{center}
\caption{Volume of chaotic phase space (positive Lyapunov exponent) over total phase space (lines are guides to the eye). 
Left $v_{22}=0.05$, right $v_{22}=0.25$.}
\label{fig:Ratio}
\end{figure}

Fourth, the width of the Gaussian (variance) diminishes with increasing energy. Fifth, with increasing energy, a strong linear correlation develops between the expectation value of $\lambda$ and energy $E$, which can be represented by a linear fit $<\lambda> = \lambda^{(0)} + \epsilon E$. 
(fit parameters are: $\lambda^{(0)}_{cl} = -0.040$, $\epsilon_{cl}=0.033$ and   $\lambda^{(0)}_{qm} = -0.026$, $\epsilon_{qm}=0.024$, i.e. 
$\epsilon_{qm} < \epsilon_{cl}$).
All those features hold for both, the classical system and the quantum system. 
A quantitative difference between classical and quantum system is apparent in the degree of chaoticity $R$, shown in Figs.[\ref{fig:Zero_Lyap_4xE_L25},\ref{fig:Pos_Lyap_4xE_L25}]. 
The corresponding results for $v_{22}=0.05$ are given in Figs.[\ref{fig:Zero_Lyap_4xE_L05},\ref{fig:Pos_Lyap_4xE_L05}], which show qualitatively the same behavior. The basic difference between the data for $v_{22}=0.25$ and $v_{22}=0.05$ shows up in the degree of chaoticity $R$ which is smaller for the latter, and also in the location of the peak in the distribution of positive Lyapunov expenents (compare Figs.[\ref{fig:Pos_Lyap_4xE_L25},\ref{fig:Pos_Lyap_4xE_L05}]), which is also smaller for the latter. 
This shows that the parameter $v_{22}$ which controls the degree of chaoticity in the classical system, has a counterpart in the parameter $\tilde{v}_{22}$ in the quantum system, which plays a similar role. 
For all energies, and all values of $v_{22}$, $R$ plotted in Fig.[\ref{fig:Ratio}] is smaller in the quantum system than in the corresponding classical system, i.e.
the classical system turns out to be more chaotic than the quantum system. This is the main result of this work.

How can we understand the observed behavior? 
The distribution of Lyapunov exponents approaching a Gaussian with increasing energy is well known from Hamiltonian systems with mixed dynamics \cite{Prasad}. The shape of the curve and its decreasing width can be understood as a consequence of the central limit theorem, which predicts such behavior for an arithmetic mean of $N$ random variables for $N \to \infty$. On the other hand, regular islands in an otherwise chaotic phase space are known to give non-Gaussian tails. What can we say about those tails?
In Hamiltonian chaotic systems, when increasing the energy then phase space becomes more chaotic, i.e. regular islands become smaller or disappear.
Thus it is plausible that the non-Gaussian tails of the distribution of positive Lyaponov exponents, associated with the regular islands, diminish.
We have estimated numerically the weight of non-Gaussian tails using a fit of the data by a Gaussian distribution. For example, in the distribution of classical Lyapunov exponents corresponding to $v_{22}=0.25$ (see left column of Fig.[\ref{fig:Pos_Lyap_4xE_L25}]), the relative error $\epsilon$ of such fits is as follows: $E=2, ~ \epsilon=0.60$, $E=4, ~ \epsilon=0.39$, 
$E=6, ~ \epsilon=0.30$, $E=8, ~ \epsilon=0.27$, which diminishes with increasing energy. Qualitatively similar behavior is found in the corresponding quantum Lyapunov exponents as well as for the data corresponding the coupling parameter $v_{22}=0.05$.

Second, quantum fluctuations which occur in the parameters of the quantum action can be viewed as renormalisation effects of the classical parameters \cite{Q5}. The value of $v_{2}$ (if sole term of potential would render the system integrable) is increased, while $v_{22}$ (term which drives the system away from integrability) is decreased (see Tab.[\ref{tab:ParamAction}]. Hence chaos is reduced in the quantum system. Such increase of the quadratic term has been observed in other systems also and seems to occur more generally.

\section{Conclusion}
\label{sec:Conclusion}
What is the significance of our findings? In particular, what is the physical significance of chaotic properties of trajectories for the quantum system?
(i) We found that the quantum action for large transition times allows to construct a quantum analogue phase space for Hamiltonian systems with mixed chaotic dynamics. Such phase space resembles the classical phase space, in particular, when the parameter driving chaos is small. For the mixed system considered, the quantum analogue phase space, like classical phase space, shows islands of regular behavior interwoven with chaotic regions. Moreover, hyperbolic fixed points occuring in classical phase space survive in the quantum phase space. 
We compared quantitatively the classical system with the quantum system and found that the latter is systematically less chaotic than the former. The reason for such behavior, in our opinion, is the tendency of quantum fluctuations to drive the quantum action nearer to a Gaussian fixed point.
Moreover, we have investigated the quantum analogue phase space as a function of energy. We find that such phase space, like the classical phase space, becomes more chaotic when energy increases. Thus introducing a quantum analogue phase space gives a detailed and quantitative measure reflecting how much chaos grows as a function of energy.
 
(ii) The significance of trajectories can be seen by drawing an analogy with 
the interpretation of scattering experiments. In a scattering experiment, one usually measures counting rates in detectors, which when positioned at different angles give a differential cross section. Using theoretical input, like the S-matrix, unitarity, and symmetries like rotational symmetry of the potential, 
one can obtain derived quantities, like scattering phase shifts $\delta_{l}(E)$.
They give a picture of scattering as function of energy and indicate the existence of resonances in certain channels (jumps of phase shift).  
The partial wave expansion in conjunction with phase shifts $\delta_{l}(E)$ 
gives a representation of the S-matrix and of scattering cross sections. 
In analogy to this picture, we consider the quantum action and its trajectories as a parametrisation of the quantum mechanical transition amplitudes. The quantum action and its trajectories are derived quantities. These trajectories 
in conjunction with tools of nonlinear dynamics (Poincar\'e sections, Lyapunov exponents) give a picture of (makes "visible") the chaotic dynamics in the quantum system. One may view the Lyapunov exponents computed from trajectories of the quantum action in analogy to scattering phase shifts computed from a partial wave analysis of scattering cross sections. 

(iii) The trajectories of the quantum analogue phase space gain a particular importance in the interpretation of of the phenomenon of chaos assisted quantum dynamical tunneling. This phenomenon is similar to tunneling through a potential barrier, however, in this case the classical transition is forbidden by a dynamical law, the presence of a KAM surface.
The experiments with ultra cold atoms in an amplitude modulated optical standing wave reported by Hensinger et al. \cite{Hensinger} 
and Steck et al. \cite{Raizen} demonstrate that such dynamical tunneling exists and its amplitude is enhanced by chaos. The tunneling occurs between regions which in classical phase space correspond to regions of stable motion (fixed points). In the quantum system this region is associated with eigenstates of the Floquet operator. Hensinger et al. point out that "Floquet states do not necessarily overlap well with the regions of regular motion of the classical Poincar\'e map, except in the semi-classical limit". 
Steck et al. work with wave packets narrow in momentum space but not localized in position. In order to get a picture of phase space they use a classical phase space with a distribution with the same position and momentum marginal distribution as the Wigner function. We suggest that in order to better understand the mechanism of dynamical tunneling, the role of localized states, decoherence, and a higher number of Floquet states, a quantum analogue phase space and its trajectories from the quantum action are a complementary tool and should give additional useful insight in the dynamics of the quantum system.

\vspace{0.5cm}
\noindent {\bf Acknowledgements} \\ 
H.K. and K.M. are grateful for support by NSERC Canada. G.M. and D.H. have been supported by FCAR Qu\'ebec. 
For discussions and constructive suggestions H.K. is very grateful to A. Okopinska and B. Eckhardt.

\end{document}